\documentclass[epj]{svjour}
\usepackage{graphics}
\def\lsim{\raise0.3ex\hbox{$<$\kern-0.75em\raise-1.1ex\hbox{$\sim$}}}
\def\gsim{\raise0.3ex\hbox{$>$\kern-0.75em\raise-1.1ex\hbox{$\sim$}}}
\begin{document}
\title{Heavy Quark Interactions in Finite Temperature QCD}
\author{S. Digal\inst{1}, O. Kaczmarek\inst{2}, F. Karsch\inst{2,3}
 and H. Satz\inst{2,4}
}                     
%
%
\institute{\inst{1} Department of Physics, University of Tokyo, Tokyo
113-0033, Japan;\\
\inst{2} Fakult\"at f\"ur Physik, Universit\"at Bielefeld,
D-33615 Bielefeld, Germany; \\
\inst{3} Physics Department, Brookhaven National Laboratory, Upton, NY 11973, 
USA; \\
\inst{4} CFIF, Instituto Superior T\'ecnico, 1049-001 Lisboa, Portugal}

\date{Received: date / Revised version: date}

\abstract{We study the free energy of a heavy quark-antiquark pair in a 
thermal medium. We constuct a simple ansatz for the free energy for
two quark flavors motivated by the Debye-H\"uckel theory of screening. 
} 
\maketitle
\section{Introduction}
\label{intro}
The theory of quantum chromodynamics (QCD) predicts that at high temperature
and density there is phase transition from hadron gas to quark gluon plasma 
(QGP). This QGP phase is believed to be formed in the collision of very
high energy heavy-ions. Various probes have been proposed to detect the 
formation of QGP in high energy heavy-ion collision experiments. One of the
clean signatures of QGP formation is the suppression of the $J/\Psi$ 
\cite{satz}. At high temperatures and density, medium effects screen 
the interaction 
between the heavy quark and antiquark pairs $(\bar{Q}Q)$. The Debye screening 
radius characterizes the distance beyond which significant thermal 
modification arise in the heavy quark potential. 
Qualitatively, one may expect that
the $\bar{Q}Q$ dissociates in the medium when the Debye radius
becomes smaller than the radius of $\bar{Q}Q$ bound state.
Quantitative predictions, such as 
dissociation temperatures or energy densities, however, require the study of 
quarkonia in a medium. Recently there have been some finite temperature
studies of $\bar{Q}Q$ spectra in quenched lattice QCD \cite{saumen}.
Previously, the medium dependence of quarkonia had been studied in the 
Schr\"odinger equation formalism. For these calculations, one needs the 
potential between the heavy quark and antiquark in the medium. However, this 
potential is not yet well understood. Only for very small distances or 
very high temperatures one can calculate it perturbatively. On the other 
hand, the free energy $F(r,T)$ of a static quark-antiquark pair can be 
calculated in lattice QCD, and using thermodynamic relations, 
one can then derive the change in energy of a thermal medium due to the
presence of a $\bar{Q}Q$ pair from this free energy. This may be more
closely related to the  desired potential. In any case, it is 
desirable to understand how the free energy $F(r,T)$ depends on $r$ and $T$. 
So far, a functional form of $F(r,T)$ in accord with the lattice data over 
wide range of temperatures is still missing. In this work we construct such a 
form of $F(r,T)$, based on studies of screening in the Debye-H\"uckel 
theory\cite{dixit}. Although there are only two parameters in our analytical 
form of $F(r,T)$, we can fit the lattice data quite well for all $r$ and for 
a wide range of temperatures, $0\le T\le 2~T_c$.

This paper is organized as follows. In section 2 we discuss the
screening of the free energies of static quark anti-quark pair in the 
presence of medium and derive our ansatz for
$F(r,T)$. In section 3 we fit our form
of free energy $F(r,T)$ to the lattice results, thus obtaining the temperature 
dependence of the free parameters in $F(r,T)$. Section 4 contains
discussions and conclusions.

\section{Medium effects on the interaction of $\bar{Q}Q$}

The free energy of a static quark-antiquark at zero temperature is given by
\cite{cornell}
\begin{eqnarray}
F(r,T=0) = \sigma r - {\alpha \over r}, 
\label{Fcornell}
\end{eqnarray}
where $\sigma$ is the string tension and $\alpha$ is the gauge coupling 
constant. The first term is due to formation of flux tube or string between 
the $Q$ and the $\bar{Q}$ when they are pulled apart. 
This string free energy increases linearly with the separation of
the $\bar{Q}Q$. The second term in the above equation corresponds to the
Coulomb interaction between the static charges. This free energy
$F(r,T=0)$ describes the charmonium and bottomonium spectrum quite well 
in terms of $\alpha$ and $\sigma$. 

The free energy $F(r,T)$ increases linearly with $r$ until at
some $r_0$, it becomes energetically favorable for the string
to break into two heavy-light mesons $(D,B)$ rather than stretching
further. After the string breaks at $r_0$, the free energy remains
constant for all separations $r$ larger than $r_0$. The value the free
energy $F(r_0,T=0)$ at the string breaking point $r_0$ can be calculated.
Thus, the masses of the $J/\Psi$, the $D$ and $F(r_0,T=0)$
are related by 
\begin{eqnarray}
2M_D &=& M_{J/\Psi} + {\rm EB} \nonumber \\
     &=& 2m_c + F(r_0,T=0).
\end{eqnarray}
\noindent Here EB denotes the binding energy and $M_D$, $M_{J/\Psi}$ and 
$m_c$ the masses of the $D$, the $J/\Psi$ and the charm quark, respectively. 
One can write a similar equation for the bottomonium system. Rearranging 
the terms, one arrives at the following relation
\begin{eqnarray}
F(r_0,T=0) \simeq 2(M_D - m_c) \simeq 2(M_B - m_b).
\end{eqnarray}
Assuming universal behavior of the string one will get a universal value of 
$r_0$, as both $2(M_D - m_c)$ and $2(M_B - m_b)$ are roughly equal to $1.2$ GeV.
Using the conventional value of $\sigma \simeq (0.4~ GeV)^2$
for the string tension, we get

\begin{eqnarray}
r_0 \simeq {2(M_{D,B}-m_{c,b}) \over \sigma}\simeq
{1.2 {\rm GeV} \over \sigma} \simeq 1.5 {\rm fm} \quad . \nonumber
\end{eqnarray}

At finite temperature, medium effects screen the heavy quark-antiquark 
interactions. The mechanism is quite well understood for Coulombic
systems and leads to the well-known Debye-screened Coulomb potential
and the corresponding Coulombic free energies \cite{Landau}. 
In general we may separate the free energy $F(r,T)$ of a pair of static 
charges into contributions arising from its potential energy at zero 
temperature $F(r,T=0)$, and a screening function $f(r,T)$,
\begin{equation}
F(r,T) = F(r,0) f(r,T)\quad .
\label{generic}
\end{equation}
This can be formulated in
the framework of Debye-H\"uckel theory \cite{Landau} and can easily
be solved for Coulombic systems. The Debye-H\"uckel ansatz
has been generalized to a large class of
potentials in arbitrary dimensions \cite{dixit}. In the following we 
briefly discuss the formalism of the Debye-H\"uckel theory and will 
then come back to the problem of finding an appropriate screening ansatz 
for the heavy quark potential.

\subsection{Debye-H\"uckel Theory}

In the framework of Debye-H\"uckel theory, the free 
energy $F(r,T)$  can be derived from a corresponding Poisson equation
\cite{dixit}. For example, a free energy $F$ at $T=0$ with the form,
\begin{eqnarray}
F = \beta r^\eta 
\end{eqnarray}
\noindent in three space
dimensions can be obtained by solving the Poisson equation
\begin{eqnarray}
-{\nabla ^2 F \over r^{\eta+1} } +
{(\eta+1) \over r^{\eta+2}} \vec{\nabla}F\cdot\hat{r} = 
4\pi\beta \delta(r).
\end{eqnarray}
In the medium, the interaction between the heavy quark-antiquark changes
because the heavy quarks polarize the medium. In linear response theory
this leads to the following substitution for the source/charge term in the 
Poisson equation
\begin{eqnarray}
\delta(r) \longrightarrow \delta(r) + A F.
\end{eqnarray}
All effects of the medium are contained in the factor $A$. 
In the case of abelian charges, $A$ is function of the density of charges in 
the medium. In a non-abelian medium, in which we are interested, making 
an estimate of the density of color charges is very difficult. However, if 
the free energy is known, one can give an estimate of $A$. This is what we 
basically do in this work. $A$ has the mass dimension of $2$ in 3-dimensional
physical space. Defining the screening mass $\mu = (4\pi\beta A)^{1/(\eta+3)}$,
one arrives at the following modified Poisson equation,
\begin{eqnarray}
{1 \over r^{\eta+1} } {d^2F \over dr^2} + { 1-\eta \over  r^{\eta+2}} {dF 
\over dr} - \mu^{\eta + 3} F = -4\pi\beta \delta(r) \quad.
\end{eqnarray}
The solution of this equation gives the free energy of static charges
in a thermal medium interacting through a potential described
by equation~(5). The equations in terms of screening 
functions can be obtained by substituting equation~(\ref{generic}) in the 
above equations. In the following we consider these separately for the case of 
string and Coulomb free energies, by choosing the appropriate 
values of $\beta, \eta$ in equation (5). 

\subsection{Coulomb Screening}

In this section we obtain the $r$ and $T$ dependence of the Coulomb 
screening function $f_c$. The temperature dependence appears only through
the screening mass $\mu$, as one can see from equation (8). For Coulomb
interactions, the two parameters of equation (5) are
$\beta=-\alpha,~ \eta=-1$. Substituting the form 
$F_c(r,T)= -(\alpha/r)f_c(r,T)$ for the Coulomb free energy in equation (8),
we get the following equation 
\begin{eqnarray}
{1 \over r}{d^2f_c \over dr^2} - {\mu^{2} \over r}f_c = -
4\pi \delta(r).
\end{eqnarray}
for $f_c(r,T)$.
It is easy to see from this equation that $f(r,\mu(T))$ will depend
only on the combination $\mu r$. The solution is given in terms
of the modified Bessel function $K_{1/2}$ and becomes 
by,
\begin{eqnarray}
f_c &=& {2^{1/2} \over {\Gamma[1/2]}} \sqrt{\mu r}
K_{1\over 2}\left(\mu r\right)\nonumber \\
&=&e^{-\mu r},
\end{eqnarray}
\noindent which is in the familiar Debye screening form. However, in the
presence of medium the Coulomb free energy has to satisfy an additional
boundary condition when the static charges are infinitely far apart;
it must then  
approach the value $F(r=\infty,T)=-\alpha\mu$ for each pair of opposite 
color charges \cite{Landau,jengo}. Taking this fact into account, the Coulomb 
screening function is given by
\begin{eqnarray}
f_c =e^{-\mu r} + \mu r.
\end{eqnarray}
This form satisfies all the required boundary conditions. The Coulomb
free energy in the presence of medium is thus given by
\begin{eqnarray}
F_c(r,T) = -{\alpha \over r}\left[e^{-\mu r} + \mu r\right].
\end{eqnarray}

\subsection{String Screening}

Next we will derive the string screening function. As in the Coulomb
screening case, the temperature dependence enters in the free energy
only through the screening mass $\mu$. For the string interaction,
the parameter $\beta$ of equation (5) is the string tension $\sigma$,
and $\eta = 1$. We substitute the form 
$F_s(r,T)=\sigma r f_s(r,T)$ for the string free energy in equation (8),
which gives the following Poisson equation for $f_s(r,T)$,
\begin{eqnarray}
{1 \over r}{d^2f_s \over dr^2} + {2 \over r^2}{df_s\over dr} - 
{\mu^{4} r}f_s = 4\pi\delta(r).
\end{eqnarray}
Here, as in the case of equation (9), the $r$ and $\mu$ dependence
are such that the solution $f_s$ is a function of the product $\mu r$ only. 
The solution of equation (13), which satisfies the boundary condition
$f(r,\mu=0) = 1$, is given by
\begin{eqnarray}
f_s(x)={1 \over x}\left[ {\Gamma\left(1/4\right) \over 2^{3/2}
\Gamma\left(3/4\right)}-{\sqrt{x} \over 2^{3/4}\Gamma\left(3/4\right)}
K_{1/4}\left(x^2\right)\right],
\end{eqnarray}
\noindent with $x=\mu r$, and where $K_{1/4}$ is the modified Bessel
function. The string free energy is now given by
\begin{eqnarray}
\hspace{-0.4cm}F_s(r,T) = {\sigma \over \mu}
\left[ {\Gamma\left(1/4\right) \over 2^{3/2}
\Gamma\left(3/4\right)}-{\sqrt{x} \over 2^{3/4}\Gamma\left(3/4\right)}
K_{1/4}\left(x^2\right)\right].
\end{eqnarray}
Because of the medium effects, $F_s(r,T)$ eventually stops rising linearly 
with $r$ and flattens off for large $r$, giving
\begin{eqnarray}
F_s(r=\infty, \mu) = {\sigma \over \mu} {\Gamma\left(1/4\right) \over 2^{3/2}
\Gamma\left(3/4\right)} \quad . 
\end{eqnarray}
It is important to note that the treatment of the medium effects in equation
(7) is an approximation. In principle, there can be higher order corrections
to equation (7) if the medium is not dilute enough. Also one may expect
non-trivial confinement effects on equation (7). As a simple ansatz, we here
assume that all such effects only change the argument of the modified Bessel 
function $K_{1/4}$,
\begin{eqnarray}
K_{1/4}\left(x^2\right) \rightarrow K_{1/4}\left(x^2 + \kappa x^4\right)
\quad .
\end{eqnarray}
The resulting 
higher order corrections affect the solution in the intermediate range
of $x=\mu r$. The string free energy in the medium now takes the
form,
\begin{eqnarray}
F(r,T)&=&{\sigma \over \mu}\left[{\Gamma(1/4)\over 2^{3/2} \Gamma(3/4)} -
{\sqrt{x}\over 2^{3/4}\Gamma(3/4)} K_{1/4}(x^2 + \kappa x^4)\right].\nonumber\\
\end{eqnarray}

\subsection{A Model for Screening of the \boldmath $\bar{Q}Q$ Free Energy}

In the above two subsections we have discussed separately the screening of 
the two terms contributing to the heavy quark potential given in 
equation~\ref{Fcornell}. In general, we should now solve the Poisson 
equation arising in the Debye-H\"uckel theory for the combined potential.
This, however, seems to be difficult to achieve in closed form. At
present, we therefore explore a screening ansatz for the heavy quark free
energies based on the asymptotic forms discussed above, {\it i.e.},
we assume that both terms are modified by their appropriate screening
functions.
 
This leads to the modification of the free energy $F(r,T)$ of the form
\begin{eqnarray}
F(r,T) = \sigma r f_s(r,T) - {\alpha \over r} f_c(r,T).
\end{eqnarray}
\noindent This free energy is same as the zero temperature form, 
modified by the Coulomb and string screening functions, 
respectively. These screening functions satisfy the
boundary conditions
\begin{eqnarray}
f_s(r,T)&=&f_c(r,T)= 1, ~~~~~~~~r \rightarrow 0.\nonumber\\
f_s(r,T)&=&f_c(r,T)= 1, ~~~~~~~~T \rightarrow 0.
\end{eqnarray}
\noindent The first condition arises because for distances $r << 1/T$ the 
$\bar{Q}Q$ cannot see the medium, the second because there is no 
medium at $T=0$.

The total free energy in the medium is then given by the sum of the
screened Coulomb and string free energies. Using equation (12) and (18) we
get
\begin{eqnarray}
F(r,T)&=&{\sigma \over \mu}\left[{\Gamma(1/4)\over 2^{3/2} \Gamma(3/4)} - 
{\sqrt{x}\over 2^{3/4}\Gamma(3/4)} K_{1/4}(x^2 + \kappa x^4)\right]\nonumber\\
&-&{\alpha \over r}\left[e^{-x} + x\right] \quad .
\label{free}
\end{eqnarray}

In the following we will compare this ansatz with lattice results
for the free energies of a static quark anti-quark pair in a thermal 
heat bath of quarks and gluons.

\section{Numerical Results}
The free energy $F(r,T)$ given in equation~(\ref{free})
contains only two temperature dependent parameters, 
$\mu$ and $\kappa$; we assume the string tension $\sigma$ and the gauge 
coupling $\alpha$ to be temperature-independent. Since we are
not analyzing the heavy quark free energies at very short distances, where
the running of the coupling becomes relevant, we use for $\alpha$ the
string model value $\alpha = \pi/12$.  
To determine the temperature dependence of $\mu$ and $\kappa$,
we fit $F(r,T)$ to the lattice results obtained in 2-flavor QCD
\cite{olaf}. At $r=\infty$, the 
free energy $F(r,T)$ depends only on $\mu$ and not on $\kappa$, 
\begin{eqnarray}
F(T)= {\sigma \over \mu (T)} {\Gamma\left(1/4\right) \over
2^{3/2} \Gamma\left(3/4\right)} - \alpha \mu(T) \quad, 
\end{eqnarray}
with $F(r=\infty,T) = F(T)$. To obtain the temperature dependence of $\mu$,
we fit this $F(T)$ to the lattice data for $r=\infty$. For $\mu(T)$, 
this gives the following form as function of $F(T)$, 
\begin{eqnarray}
\mu(T)&=& - {F(T) \over 2\alpha}\nonumber\\
&+&
\sqrt{\left(F^2(T) + 4\sigma{\Gamma\left(1/4\right) \over 2^{3/2}
\Gamma\left(3/4\right)}\right)/4\alpha^2} \quad.
\end{eqnarray}
In Fig.\ 1 we show the lattice results for $F(T)$, obtained from
the mentioned lattice calculations in 2-flavor QCD \cite{olaf}, as well as
the corresponding fit results for the screening mass $\mu(T)$. 

\begin{figure}
\resizebox{0.40\textwidth}{!}{%
  \includegraphics{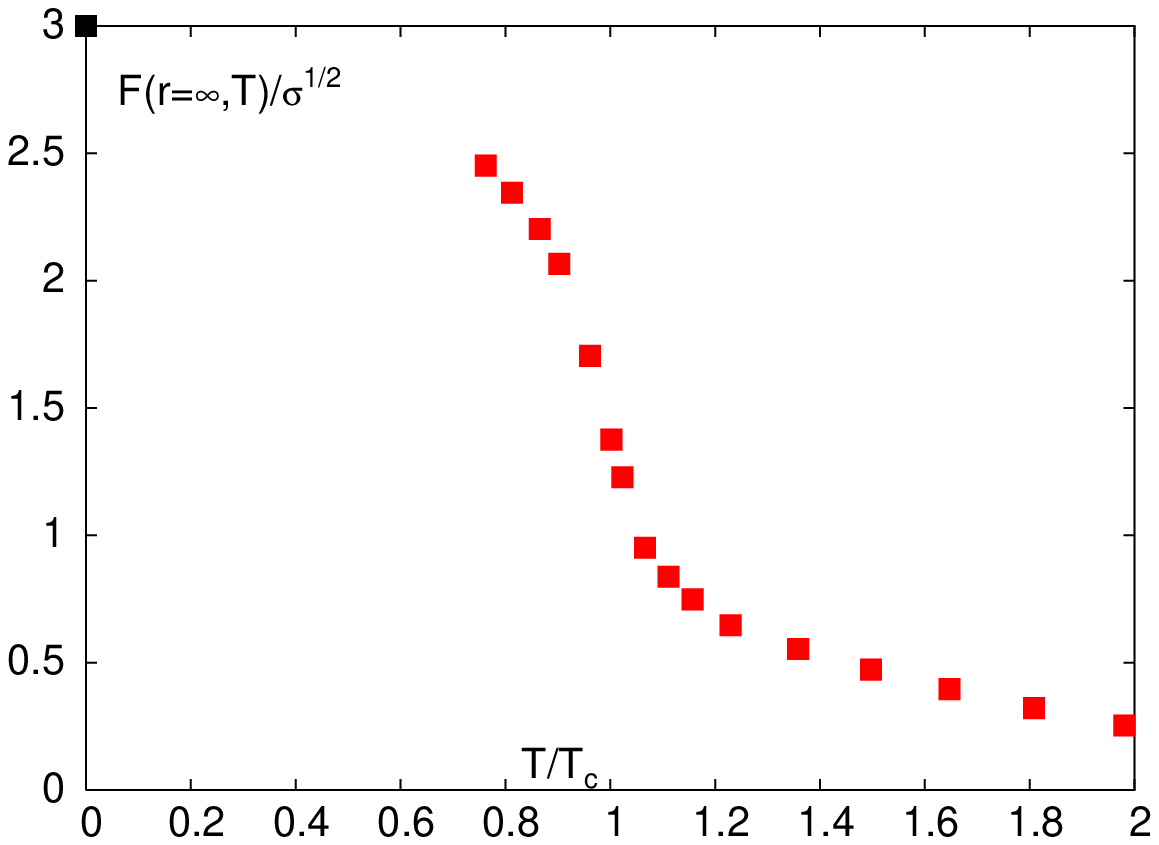}
}
\vspace*{-0.1cm}
\resizebox{0.40\textwidth}{!}{%
  \includegraphics{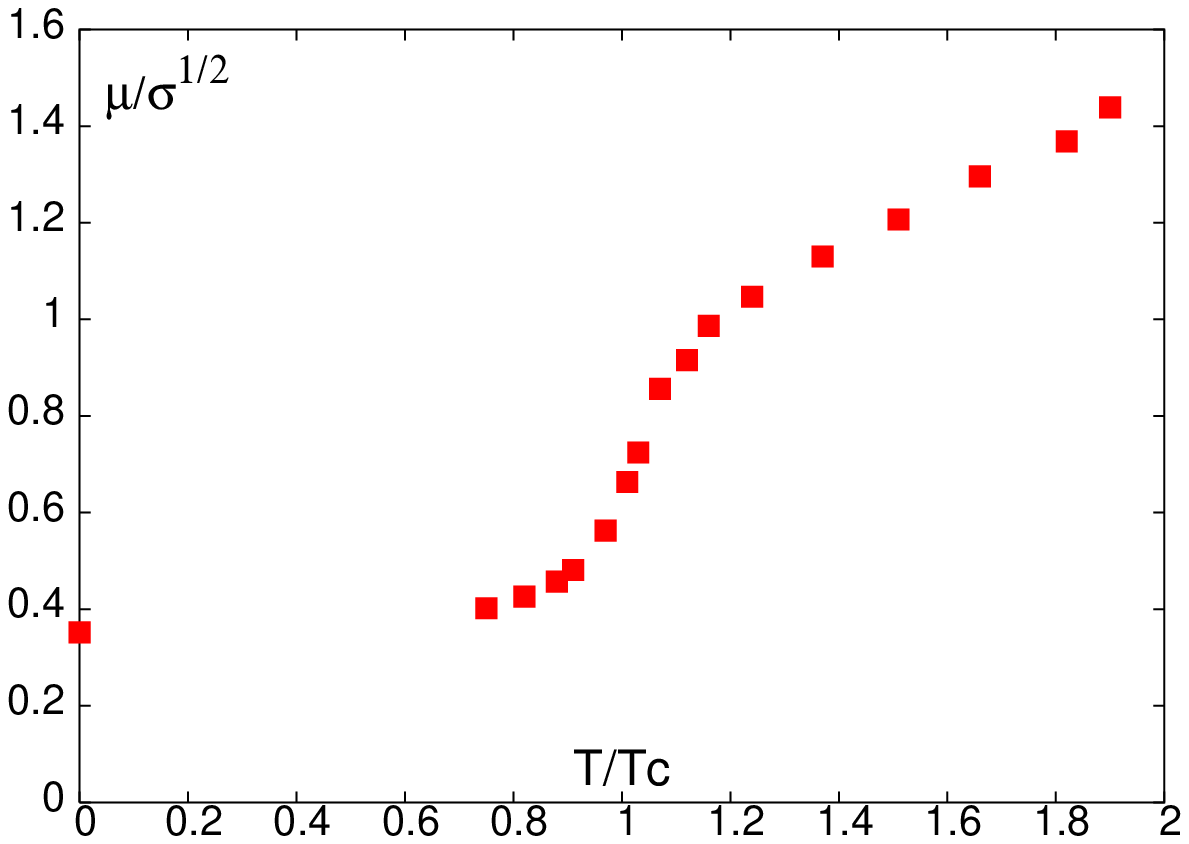}
}
\caption{Results from 2-flavor lattice QCD calculations 
for the large distance limit of the free energy (upper figure) and the 
corresponding screening mass (lower figure)}
\label{fig:1}       
\end{figure}

Once we have the temperature dependence of $\mu(T)$, we fit equation (21)
to the lattice data to obtain $\kappa(T)$. Our form of the free energy 
$F(r,T)$ fits these data quite well, as seen in Fig.\ 2, where
we show our fit curves (solid lines) together with the lattice results. 
\begin{figure}
\resizebox{0.40\textwidth}{!}{%
  \includegraphics{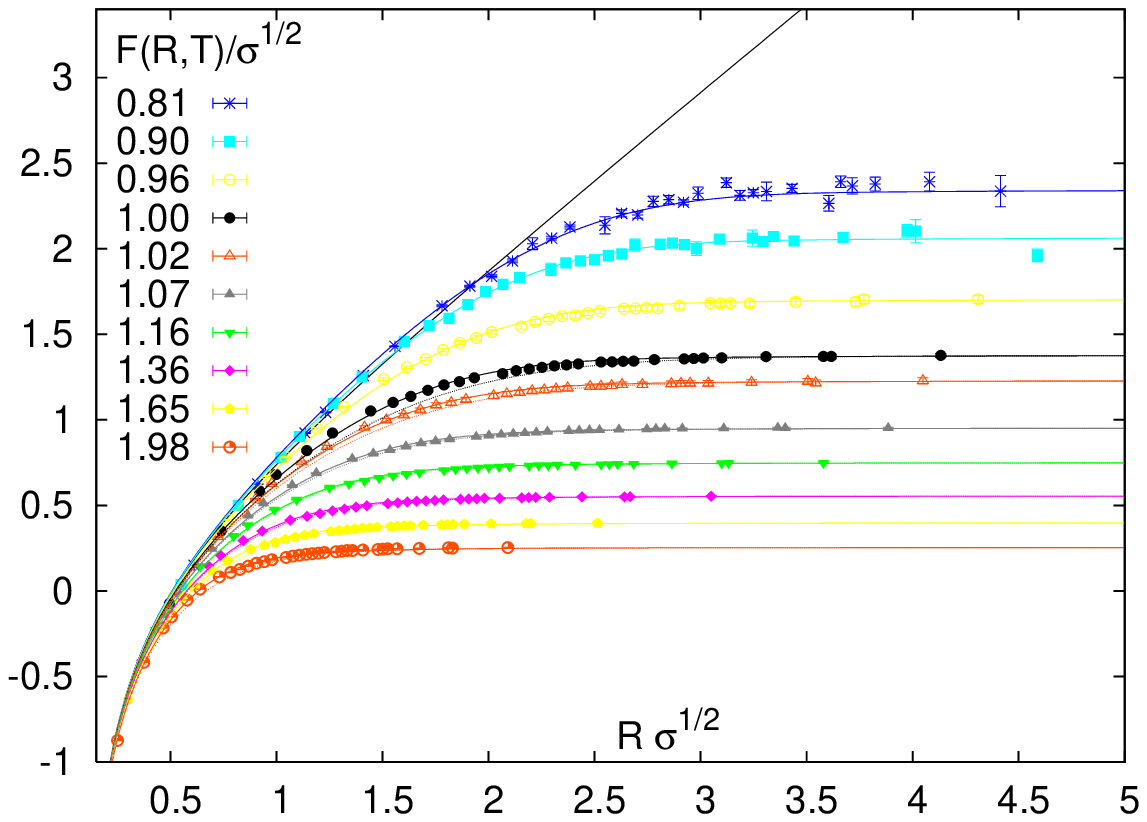}
}
\vspace*{-0.1cm}
\resizebox{0.40\textwidth}{!}{%
  \includegraphics{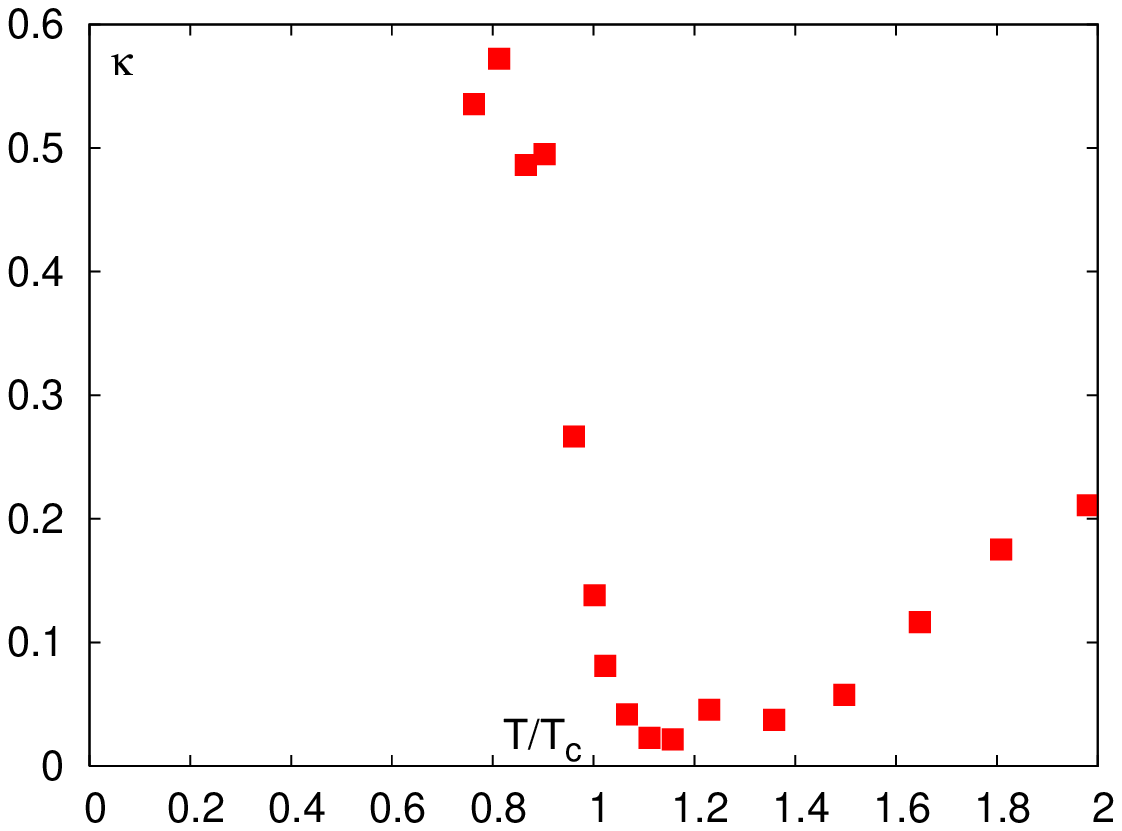}
}
\caption{Fits to the lattice results for the free energy as function of 
$r$ for different temperatures (upper figure), and for the parameter $\kappa$
(lower figure).}
\label{fig:2}       
\end{figure}
The fit results for the temperature dependence of $\kappa(T)$ are also
shown in Fig.\ 2. They indicate that $\kappa(T)$ is finite and nearly 
constant below $T\; \lsim\; 0.8 T_c$, which may reflect the effect of 
confinement and of string breaking as an additional {\it non-thermal} 
screening effect, which we have ignored here. 
Closer to $T_c$, $\kappa(T)$ drops sharply and remains small above $T_c$. 
In fact, above $T_c$ we obtain good fits also with $\kappa$ set to zero
(as shown by the lower lines for T=$T_c$ and $1.02~T_c$ in Fig.\ 3).
This suggests that our ansatz with two separate screening functions
provides a good approximation to the correct screening solution of
the Debye-H\"uckel theory in the QCD plasma phase.

\section{Discussions and Conclusions}

Using Debye-H\"uckel theory, we have constructed a simple functional 
form for the free energy $F(r,T)$ of a static 
quark-antiquark pair in a thermal heat bath. For constant string tension and
gauge coupling constant, $F(r,T)$ is specified by only two 
temperature-dependent parameters, $\mu$ and $\kappa$. We determine 
their temperature dependence by fitting our form of 
$F(r,T)$ to the lattice results for the free energy. The resulting
$F(r,T)$ fits the lattice data quite well for all $r$ and in a broad range
of temperatures from zero to $2~T_c$. Since our $F(r,T)$ is an analytic form,
one can now obtain from it the potential 
energy of a static quark anti-quark system
using the thermodynamic relations.
This may then be applied to study medium effects on the
properties of heavy quarkonia.

\end{document}